# Improving accuracy and uncertainty quantification of deep learning based quantitative MRI using Monte Carlo dropout

Mehmet Yigit Avci[1,3], Ziyu Li[2], Qiuyun Fan[3,4,5], Susie Huang[3,4], Berkin Bilgic[3,4], Qiyuan Tian[3,4,5*]

[1]Department of Informatics, Technical University of Munich, Munich, Germany;

[2]Nuffield Department of Clinical Neurosciences, University of Oxford, Oxford, UK;

[3]Athinoula A. Martinos Center for Biomedical Imaging, Department of Radiology, Massachusetts General Hospital, Charlestown, MA, USA;

[4]Harvard Medical School, Boston, USA;

[5]Department of Biomedical Engineering, College of Precision Instruments and Optoelectronics Engineering, Tianjin University, Tianjin, China.

[*]**Correspondence to**: Qiyuan Tian, Athinoula A. Martinos Center for Biomedical Imaging, 149 13th Street, Charlestown, MA, 02129, United States. E-mail address: qtian@mgh.harvard.edu.

**Abstract**

**Purpose**

To improve the accuracy and quantify uncertainty of the neural networks and reduce the data requirement for safer and more accurate deep learning based quantitative MRI.

**Methods**

A U-Net was modified to include dropout layers (active during both training and inference) for decoding to obtain high quality fractional anisotropy (FA) and mean diffusivity (MD) maps from only 3-direction MRI scans. Ground-truth FA and MD volumes were derived from all 108 volumes using FSL's "dtifit" function. Final estimated maps are obtained as the mean of 1~100 predictions. Uncertainty maps are generated which were calculated as variance over mean of 100 predictions. The accuracy was measured in terms of mean absolute errors (MAEs) between neural network estimates and ground-truth FA and MD values. Experiments were performed using different numbers of subjects for training (1~32) and dropout rates (0~0.7).

**Results**

The accuracy can be improved significantly, up to 25%, compared to a U-net without dropout, especially with a limited number of training subjects. Averaging predictions from the network with Monte Carlo dropout trained with one subject performed almost equivalently with the network without dropout trained with four subjects. The 3D structured network performed better than the 2D counterpart. Uncertainty was higher in pathology regions.

**Conclusion**

Monte Carlo dropout is a very effective tool for uncertainty estimation which also has the ability to improve the accuracy of the networks. It can be used for a wide range of networks and tasks for MRI. This work also made it possible to detect pathological regions with the use of uncertainty maps.

**Introduction**

Quantitative MRI is a highly regarded tool that has achieved widespread adoption for the measurement of soft-tissue microstructures. Among the various techniques employed, diffusion tensor imaging (DTI) has emerged as a popular method, offering valuable insights into the shape and orientation of the local microenvironment by scrutinizing the orientation of the diffusion restrictions. DTI signal interpretation entails the use of a series of specialized sequences for fitting a 3x3 tensor model, following which the tensor undergoes eigen-decomposition. The eigen-vectors extracted from the tensor provide information about the directionality of microstructure, while the eigen-values provide the basis for generating quantitative metrics such as fractional anisotropy (FA) and mean diffusivity (MD), which have accustomed in studies of neurodevelopment and neurological disorders. However, a minimum of seven inputs (six unique diffusion-encoding directions and a non-diffusion-weighted scan) is mandatory for fitting the tensor. Conventional practice dictates the use of a larger number of diffusion-encoding directions, as this enhances the precision of the fitting process.

Numerous machine learning algorithms have been proposed to mitigate the data requirements in MRI, thereby enabling the successful generation of high-quality quantitative metrics. Early attempts used multilayer perceptron networks, such as Golkov et al. estimated diffusion kurtosis measures from only 12 directions and neurite orientation dispersion and density measures from 8 directions[1] and Aliotta et al. generated FA and MD maps from raw DTI signals with only three diffusion-encoding directions[2]. The power of convolutional neural networks (CNN) is shown in some recent studies that give promising results thanks to CNN's ability to use the spatial information in surrounding voxels. Tian et al. proposed a CNN to generate high quality diffusion signals from low quality diffusion signals along with anatomical images (i.e., T1, T2) and performed DTI fitting[3], another work from Aliotta et. al. employed a 2D U-net model to directly estimate FA and MD maps from diffusion images[4], in another work, a 2D U-net model is also used for estimating T1 and T2 maps from three contrast-weighted images[5]. Furthermore, promising advancements have been made in the field by the utilization of sophisticated architectures. For instance, Sveinsson et al. produced synthetic T2 maps using anatomic MRI scans not designed for T2 mapping by leveraging the power of conditional

GANs[6], while Karimi et al. used transformers to estimate accurate diffusion tensor from six diffusion gradients [7].

While the primary goal of these works was to improve accuracy, the importance of quantifying their uncertainty or confidence levels has often been overlooked. However, such assessments are equally crucial as accuracy itself. It is essential to exercise caution when relying on the output of machine learning models, especially in medical applications where images can influence critical, life-saving decisions. While a neural network may accurately predict a case, it may perform poorly for another if its general uncertainty is high. NN's also may have varying uncertainty for estimating different quantitative metrics and/or in different regions/tissues. Moreover, generalization of NN's to test images may not be possible due to the presence of artifacts and structural pathology or different imaging protocols and hardware. Therefore, it is crucial to characterize uncertainty/confidence of NN's for risk management, robustness evaluation, and potential human intervention on failure cases.

For uncertainty quantification, recent progress in Bayesian approaches use Monte Carlo samples coming from posterior distribution. Monte Carlo dropout (MCDropout) is one of the Bayesian methods which has been used as a highly effective and scalable method. During training, dropout randomly deactivates a proportion of the neurons in a NN, thereby reducing interdependent learning and serving as a valuable regularizer to prevent overfitting. Gal and Gahramani demonstrated that incorporating dropout not just during training but also during inference can be interpreted as a Bayesian approximation of the Gaussian process[8]. Each dropout configuration corresponds to a sub-network with slightly different architecture and each forward pass yields a different prediction as a sample of the approximate posterior distribution (Fig.1a). Multiple forward passes (30-100) output a predictive distribution over the mean. Then, the uncertainty of the networks can be quantified as the standard deviation of the predictions of these multiple forward passes. Additionally, higher accuracy and reduced uncertainty can be achieved by averaging these predictions (i.e., model averaging). Dropout has been utilized in various works to quantify uncertainty with Monte Carlo samples in different tasks such as super-resolution, scene understanding, MRI image reconstruction, segmentation[9-11].

In this work, we propose to employ MCDropout for improving the performance, preventing overfitting, and quantifying the uncertainty of NN's for quantitative MRI. Our approach involves incorporating dropout layers into the decoding segment of the U-net architecture, which is commonly used NN in many MRI studies [4, 12], and entitle our network as "DUnet". High-quality FA and MD maps are synthesized from one non-diffusion-weighted image and three diffusion-weighted image (DWI) volumes- also an impossible task using DTI model fitting- along with their corresponding uncertainty maps. We also observed variations in the network's uncertainty across different regions of the brain and between FA and MD generation. Moreover, we test the capability of predictive uncertainty maps to demonstrate predictive errors which may help to detect potential failures on images of both healthy subjects and HCP subjects with benign abnormalities that are not included in the training data. Our codes are publicly available: github.com/myigitavci/dropout_ISMRM

Methods

## 2.1 Data Acquisition and Processing

Pre-processed diffusion data (1.25-mm isotropic, 18×b=0, 90×b=1000 s/mm2) of 52 healthy subjects (20 for testing) from the Human Connectome Project (HCP) were used. For Fig.5, we selected HCP subjects who have anatomical anomalies flagged with issue code A. For each subject, the first b=0 image and three DWI volumes along orthogonal directions were extracted. Ground-truth FA and MD volumes were derived from all 108 volumes using FSL's "dtifit" function. The "aseg" brain segments from FreeSurfer were resampled to the diffusion image space. For each subject, diffusion weighted images are standardized by subtracting the mean intensity and then dividing by the standard deviation of the image intensities of the brain voxels from diffusion data.

## 2.2 Network Design and Training

To evaluate the effectiveness of Monte Carlo dropout and averaging strategy in the context of quantitative MRI, we have proposed and implemented DU-net architectures based on 2D and 3D U-net models. The U-net architecture consists of an encoding part which employs max pooling operation and a decoding part which employs up-sampling operation. Skip connections were also included between the encoding and the decoding parts in order to concatenate low resolution features with high resolution features and alleviate the vanishing-gradient problem. DU-net has dropout layers on decoding parts that are active during both training and inference. The architectures are essentially a standard U-net if dropout rate is set to 0. The input layer to both models are three diffusion-weighted images and a b=0 image and the output layer predicts both FA and MD. The proposed 3D DU-net architecture is shown in Fig. 1. For both networks, input volumes are normalized within the brain mask by removing the mean value from voxel intensities and divided by their standard deviation. For 3D DU-net, due to limited GPU memory, volumetric data is separated into 64x64x64 blocks and given as inputs, for 2D U-net, inputs are given slice by slice from the whole volume. The output blocks are recombined into the brain volumes such that voxels from the block were put back to their original location in the volumetric data and the overlapping voxels were averaged. Training was performed using Adam optimizers with a learning rate of 0.0001 using Keras program interface (https://keras.io) via Tensorflow software (https://www.tensorflow.org). L2 loss was computed only within the brain parenchyma. 20% of training data is randomly separated as validation set for each epoch and the weights from last epochs with lowest validation loss-error is chosen as resulting network weights.

**2.3 Evaluation**

Experiments were performed with different numbers of subjects for training (1~32) and dropout rates (0~0.7) (the highest accuracy is achieved with 0.1~ 0.2 dropout rates, performance table is available in Supplementary Fig. 1.a) and averaging different numbers of predictions (averaging more predictions yields higher accuracy, yet the performance plateaued after 50 predictions, performance table is available in Supplementary Fig. 1.b). To generate uncertainty maps, the standard deviation over the mean of 100 predictions is calculated for each voxel. The final estimates were calculated as an average of 1~100 predictions for each voxel. Each NN-estimated FA and MD were then compared with their corresponding ground truths in terms of voxel-wise differences. The mean absolute errors (MAEs) were then computed across all brain voxels within the brain parenchyma and averaged for measuring accuracy. MD MAE's are multiplied by 1000.

**Results**

In Fig.2 two single predictions of DU-net and their differences are shown for each of MD (Fig.2 a-c) and FA (Fig.2 d-f). The predictions are slightly different from each other since dropout switches off different neurons for each case. Sample uncertainty maps of a healthy subject that are calculated from 100 predictions are also generated where it is seen that general uncertainty is lower for MD because MD estimation is a simpler task than FA estimation (Fig.2 g). Also, the uncertainty values vary across different tissues/regions. For FA, white matter regions generally have low uncertainty and gray matter regions have high uncertainty. It was found that Cerebral cortex has the highest (0.0756), corpus callosum (0.0288) has the lowest uncertainty values for FA, whereas pallidum (0.0423) has the highest, accumbens (0.0166) has the lowest uncertainty values for MD. It is also seen that there is no correlation between FA and MD uncertainties according to the regions.

Estimated FA and MD maps of DU-net trained with data of different numbers of subjects (1~32) are displayed for a representative subject along with 20-subjects-mean MAEs for networks with/without dropout and single/100-times-averaged predictions (Fig.3). Averaging 100 predictions of networks with dropout improved the estimation accuracy independent of the number of training subjects. The improvement was most significant when the number of subjects were limited since dropout avoided overfitting. Even single prediction of DU-net (dropout is on during inference) has higher accuracy than no dropout for FA, whereas single prediction for MD started to result in worse accuracy while increasing the number of subjects. Among all experiments, averaging 100 predictions of DU-net achieved the highest accuracy, and resulted in up to 24.817% lower MAE for MD and 18.143% lower MAE for FA for only one subject for training. Averaging predictions from a DU-net trained on data from one subject had almost equivalent performance (0.0595 MAE for FA and 0.0615 MAE for MD) with a prediction of a standard U-net trained on data from four subjects (0.0589 MAE for FA and 0.0650 MAE for MD). Also, improvement was more evident for small numbers of training data (1~4 subjects) for MD, and more significant for larger numbers of training data (8~32 subjects) for FA. However, networks trained with small amounts of data (1~4 subjects) yielded estimated FA and MD volumes with loss of fine details (Fig.3 a,b), and resulted in low accuracy as expected.

Moreover, results from 2D and 3D networks with/without dropout and single/100-times-averaged predictions are compared with demonstration of their residuals compared to ground-truth images for an exemplary subject Figure 4. 100-predictions-averaged results of DU-net were visually similar to the ground-truth images and quantitatively has the lowest MAE. For the 20 evaluation subjects, the group-level mean ( the group-level standard deviation) MAE result of the proposed method was 0.0414 for 3D network, and 0.0490 for 2D network for FA, and 0.0467 for 3D network and 0.0491 for 2D network for MD, whereas MAE result of single prediction of DU-net was 0.0435 for 3D network, and 0.0514 for 2D network for FA, and 0.0498 for 3D network, and 0.0498 for 2D network for MD. U-net without dropout has generally worse performance than DU-net with MAEs of 0.0460 for 3D network, and

0.0525 for 2D network for FA, and 0.0483 for 3D network, and 0.0501 for 2D network for MD. The 3D network architecture yielded significantly higher performance than the 2D counterpart for both FA and MD estimation with the proposed method.

A 3D DU-net trained only with healthy subjects is applied on HCP subjects with anatomical abnormalities. Results of three different subjects with three different abnormalities (cyst, cavernoma and calcification) are shown in Figure 5. As expected, the uncertainty in the abnormal regions is generally higher relative to the other parts of the brain, since the network is not confident with those regions that never occurred in training set. The uncertainty in the cavernoma region for the estimated FA and MD (Fig. 5 a,b) is significantly higher, and the accuracy is significantly lower compared to other regions (except CSF). Also, in the abnormality region of the subject who has a small posterior midline arachnoid cyst, uncertainty is higher, and accuracy is significantly lower for MD estimation (Fig.5c). Whereas for FA estimation (Fig.5d), although accuracy is high in pathology region, NN's uncertainty is high which alerts that there may be an abnormality or artifact. Third subject has significantly high uncertainty and low accuracy in the abnormal region which is a small calcification for MD estimation (Fig.5e) even though the abnormality is small compared to other subjects. For FA estimation (Fig.5f), accuracy is also lower, but uncertainty of the pathology region is not highlighted since CSF uncertainty is also high.

**Discussion and Conclusion**

The proposed DU-net method enabled not only higher accuracy but uncertainty quantification of NN's for estimating FA and MD maps simultaneously. It is observed that neural networks have varying uncertainty between different regions of the brain (Fig.2 j), such as white matter uncertainty is generally lower than gray matter for FA estimation. Moreover, the networks are more confident for MD estimation than FA estimation since estimating MD map is a simpler task than FA estimation in terms of structural complexity. Hereby, while evaluating the results, DU-net makes it possible to assess the confidentiality of the networks for different regions or tasks.

Even a single prediction from DU-net outperformed standard U-net prediction for FA and MD (with 1~4 training subjects). Increasing the number of training subjects allowed networks to learn more sparse and high dimensional features which resulted in more sharp and visually satisfactory images (Fig. 3). It is also worthy to note that MD volume prediction with single forward pass of DU-net is worse than no dropout U-net for 8,16,32 training subjects which yielded decrease in accuracy in percentage of 0.036, 1.049, 3.105. The reason we think is that the MD volumes are less sharp, and the voxel intensity values are more similar to each other, so employing dropout where the network has already learned from enough number of subjects only gives rise to loss of spatial information. Furthermore, 3D networks yield higher quality images as expected, because they can learn the spatial information in the 3D structures of the brain more robustly than its 2D counterpart (Fig. 4).

Also, it is shown that networks may have low confidence in their outputs if the test data is not well represented in the training data. The uncertainty in the pathology regions is significantly higher than healthy white matter and gray matter regions for MD estimation. For FA estimation, uncertainty in the pathology region is also generally higher than white matter and gray matter regions but not as obvious as MD. This may also be related to learning to estimate FA is a more complicated task than estimating MD. Networks already learn more complex structures while trying to estimate FA, so that general uncertainty is already high, and network may not spatially comprehend that the abnormality region is specifically different than a healthy tissue. Another consequence is that accuracy is also lower in abnormal regions which can be seen from the residual maps. It is also interesting that for some cases, in the abnormal regions, accuracy is high, but the uncertainty is also high, as the case in Fig.5c. These cases should be treated carefully, as for some cases pathology may not be very clear in the estimated image, but uncertainty map may indicate there is an artifact or an abnormality. Also, networks have higher uncertainty in CSF, as the networks have been trained with masks that exclude CSF, but CSF results are also shown because some abnormalities are present in CSF regions.

DU-net can be simply extended for different DTI metrics, MRI sequences and different MR imaging tasks such as reconstruction, denoising, super resolution, segmentation, classification. Also, accuracy can be improved, and general uncertainty may be reduced if more directions for DWI or different sequences such as T1w or FLAIR images are fed into neural networks.

**FIGURES**

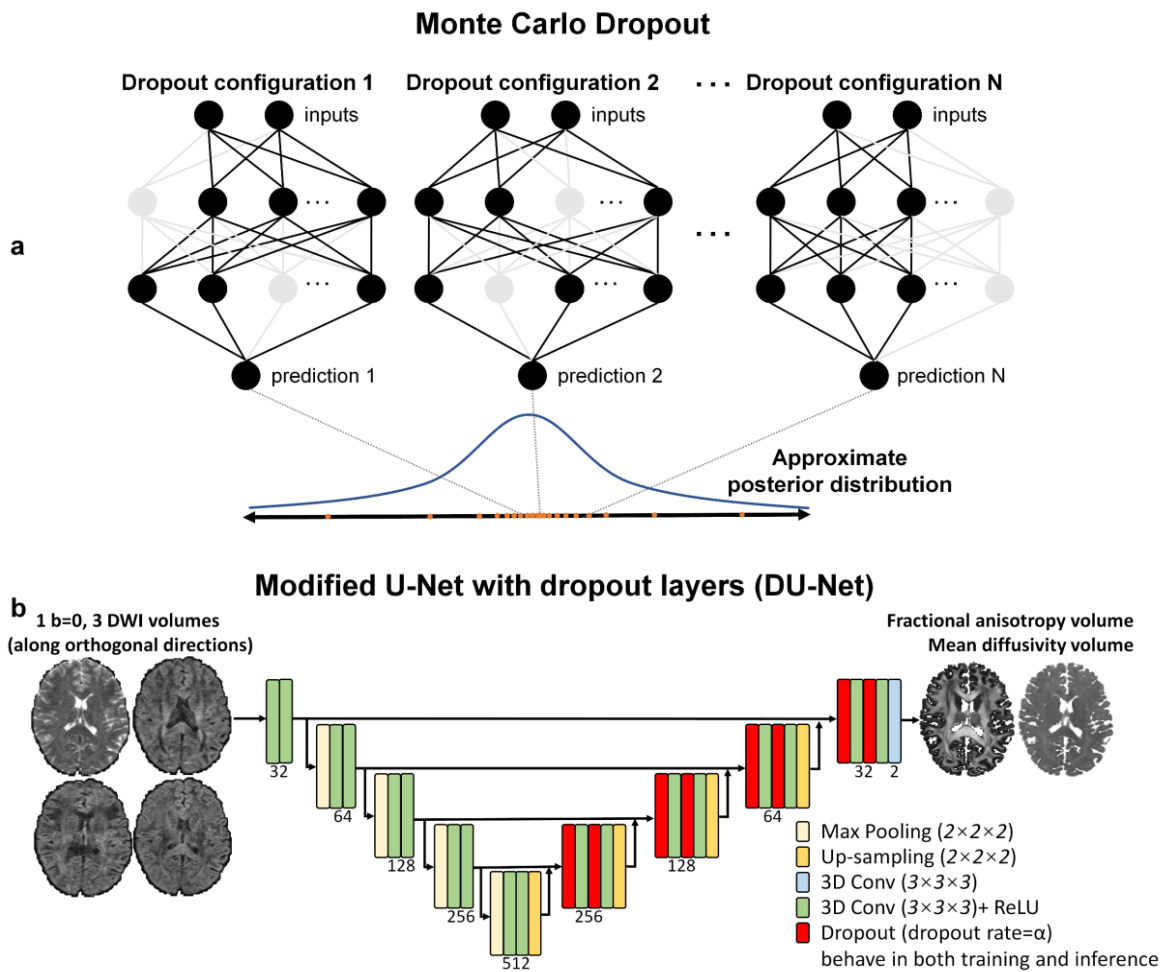

**Figure 1.** MCDropout turns on dropout during inference and generates many varying predictions as samples from approximate posterior distribution of the Gaussian process for uncertainty quantification (a). The 3D U-Net (5 depth, 32 kernels at 1st depth) is modified to include dropout layers for decoding, which are active in both training and inference (b). The input includes 1 b=0 image volume and 3 diffusion-weighted image volumes along orthogonal directions. The output includes two volumes of fractional anisotropy and mean diffusivity values from DTI.

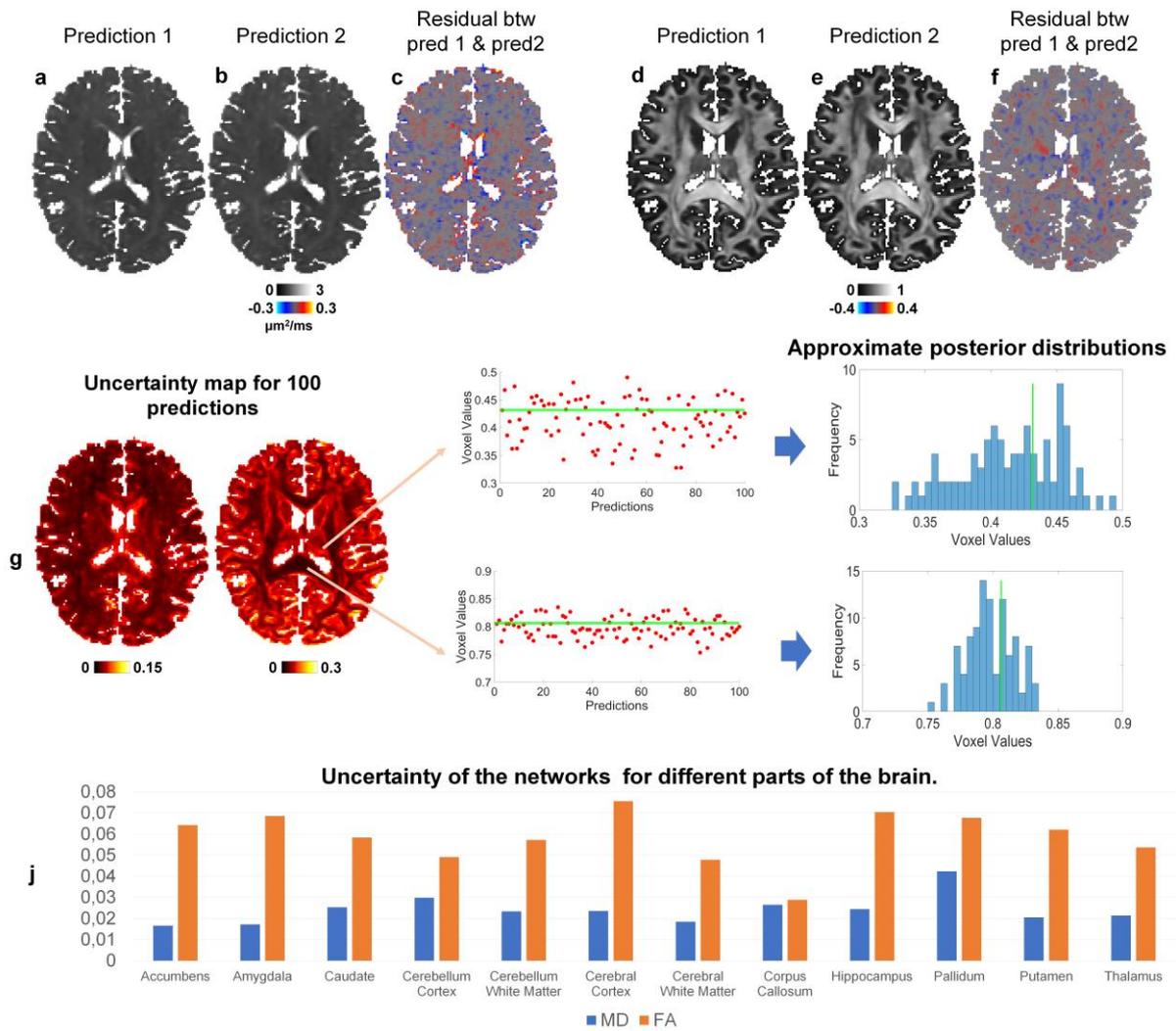

**Figure 2.** Difference between two MD predictions of DUnet (a,b,c) and difference between two FA predictions of DUnet (d,e,f). Uncertainty maps (g) calculated from 100 predictions (ground-truth values indicated by green lines) and 100 predicted values of a single voxel from white matter with low uncertainty and a voxel from gray matter with high uncertainty, along with the approximate posterior distributions (histograms) derived from them are shown. The 10-subject mean values of the averaged uncertainty within different brain structures are listed (j).

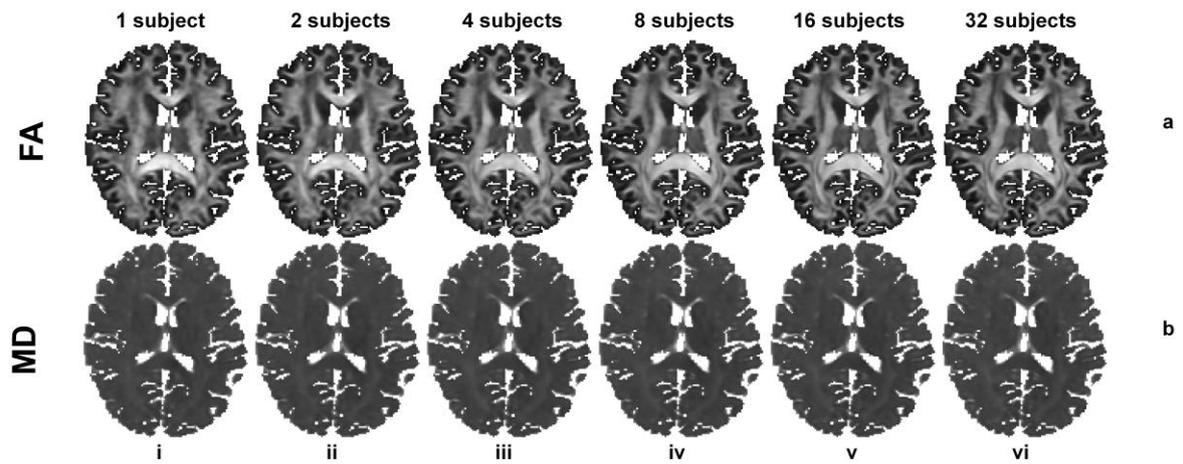

| Number of subj. for training | MAE (FA) | | | | |
|---|---|---|---|---|---|
| | W/o dropout | 1 pred. drop rate=0.2 | Improvement (%) (1 pred vs w/o dropout) | 100 preds. drop rate=0.2 | Improvement (%) (100 preds. vs w/o dropout) |
| 32 | 0.0460 | 0.0435 | 5.434 | **0.0414** | 10.000 |
| 16 | 0.0498 | 0.0460 | 7.630 | **0.0438** | 12.048 |
| 8 | 0.0525 | 0.0491 | 6.925 | **0.0469** | 10.666 |
| 4 | 0.0589 | 0.0536 | 8.999 | **0.0511** | 13.243 |
| 2 | 0.0694 | 0.0596 | 14.121 | **0.0568** | 18.155 |
| 1 | 0.0733 | 0.0622 | **15.143** | 0.0595 | 18.827 |

| Number of subj. for training | MAE (MD) (µm²/ms) | | | | |
|---|---|---|---|---|---|
| | W/o dropout | 1 pred. drop rate=0.2 | Improvement (%) (1 pred vs w/o dropout) | 100 preds. drop rate=0.2 | Improvement (%) (100 preds. vs w/o dropout) |
| 32 | 0.0483 | 0.0498 | -3.105 | **0.0467** | 3.312 |
| 16 | 0.0524 | 0.0529 | -1.049 | **0.0496** | 5.343 |
| 8 | 0.0557 | 0.0559 | -0.0360 | **0.0523** | 6.104 |
| 4 | 0.0650 | 0.0594 | 8.615 | **0.0551** | 15.231 |
| 2 | 0.0796 | 0.0662 | 16.835 | **0.0604** | 24.121 |
| 1 | 0.0818 | 0.0670 | **18.093** | 0.0615 | 24.817 |

**Figure 3.** Results of 100-prediction-averaged FA and MD estimations (a, b) for different numbers of training subjects (1-32) (i-vi). The 20-subject means of the mean absolute error (MAE) between results and ground-truth values within the brain parenchyma are listed in the table, for experiments with different numbers of training subjects. The percentage increases of group-level means from a single prediction of DU-net and 100-prediction-averaged results compared to those from standard U-Net are also listed.

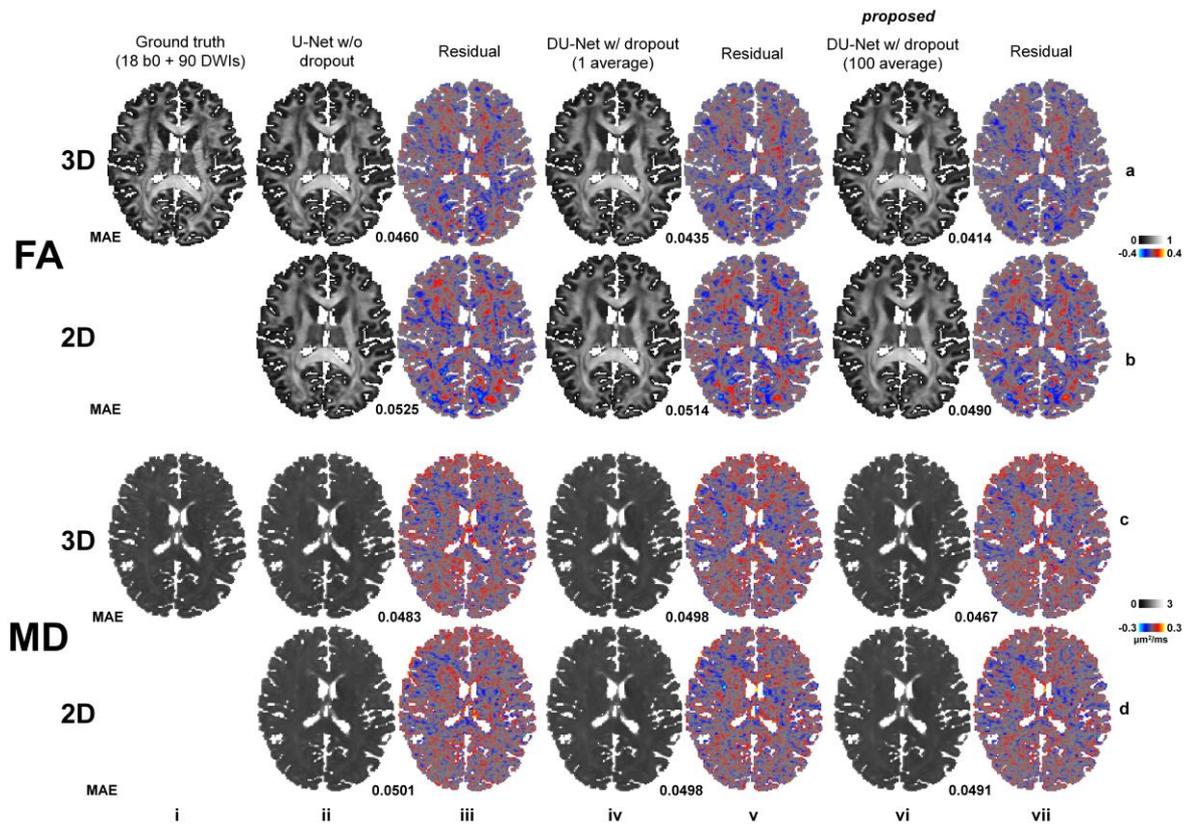

**Figure 4.** Maps of FA and MD estimates from standard U-Net (i), and a single prediction (iv) and the average of 100 predictions (vi) from DU-Net with Monte Carlo dropout for both 3D (a, c) and 2D (b, d) U-net structures, and their residuals (iii,v,vii) compared to ground-truth values are shown. MAEs below the images are 20-subject-mean results.

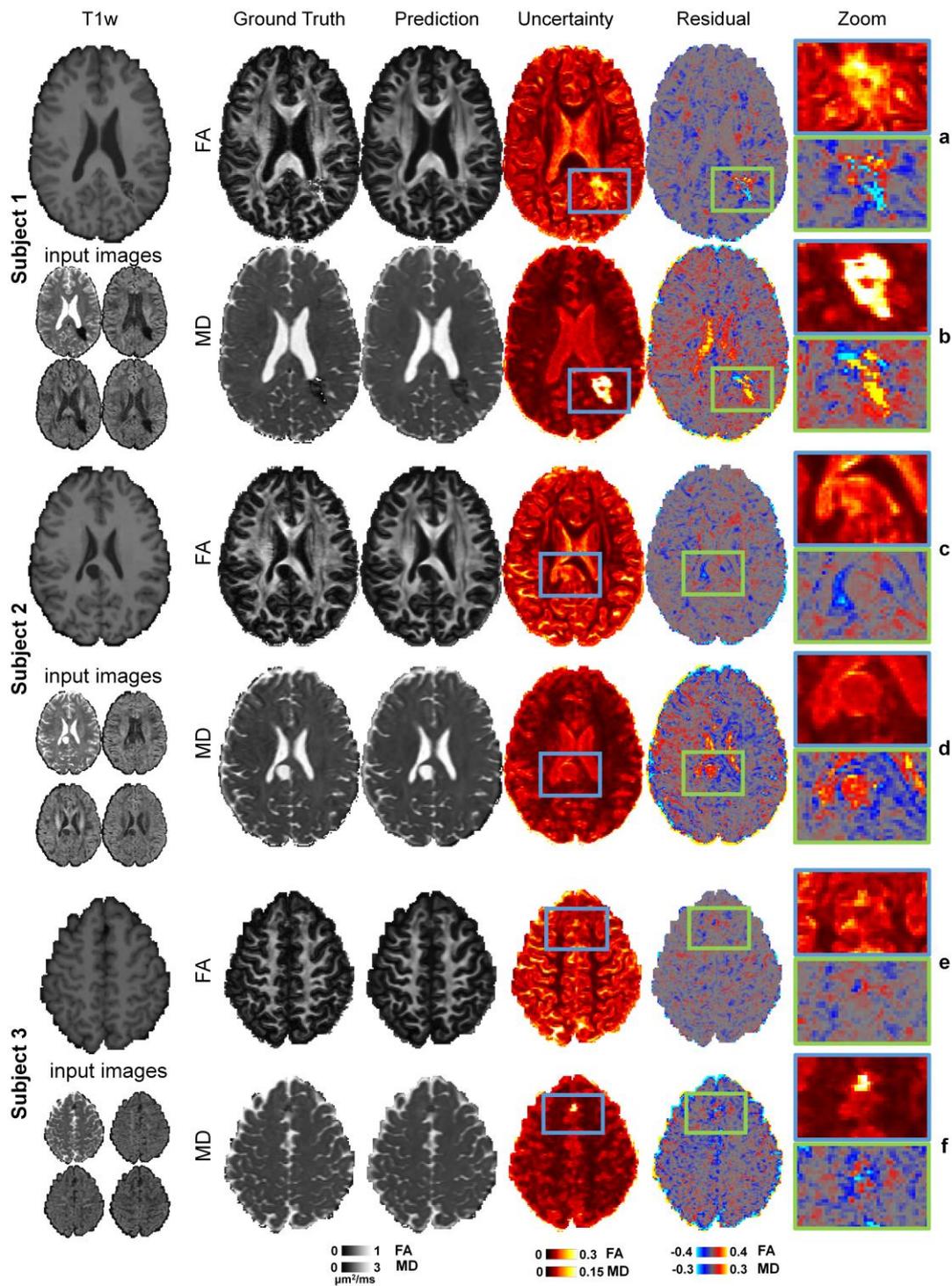

**Figure 5.** FA and MD maps and their corresponding residual against ground truth images and uncertanties of three different subjects with different abnormalities that are not present in the training data. Subject 1(a,b) has a cavernoma in left occipital lobe ,subject 2 (c,d) has a small posterior midline arachnoid cyst and subject 3 (e,f) has calcification. In the most right column, zoomed versions of residual and uncertainty images are shown.